\documentclass[preprint,showpacs,preprintnumbers,amsmath,amssymb]{revtex4}
\usepackage{graphicx}
\usepackage{amsmath}
 \usepackage{epstopdf}
\usepackage{color}

\begin{document}
\title{An experimental model of reflection and transmission of ocean waves by an ice floe}

\author{Alessandro TOFFOLI,$^1$}
\author{Alberto ALBERELLO,$^1$} 
\author{Luke G. BENNETTS,$^2$} 
\author{Michael H. MEYLAN,$^{3}$} 
\author{Claudio CAVALIERE,$^4$} 
\author{Alexander V. BABANIN$^1$}

\affiliation{%
$^1$Centre for Ocean Engineering Science and Technology, Swinburne University of Technology,
Melbourne, VIC 3122, Australia\\
E-mail: toffoli.alessandro@gmail.com\\
$^2$School of Mathematical Sciences,
University of Adelaide, Adelaide, SA 5005, Australia\\
$^3$School of Mathematical and Physical Sciences, University of Newcastle, Callaghan NSW 2308,
Australia\\
$^4$Polytechnic University of Milan,
Milan, 20133, Italy}

\date{\today}

\begin{abstract}
An experimental model of reflection and transmission of ocean waves by an ice floe is presented. 
Evolution of mechanically-generated, regular waves is monitored in front and in the lee of a solitary, square floe, made of a synthetic material.   
Results confirm dependence of reflection and transmission on the period of the incident wave. 
Results also indicate that wave overwash on the floe affects reflection and transmission.
\end{abstract}
\maketitle

\section{Introduction}

Ocean surface waves penetrate tens to hundreds of kilometres into the sea ice-covered oceans.
Wave motions force ice floes to bend and flex. 
This can cause the floes to fracture and, subsequently, to break into smaller floes. 
The region of ice-covered ocean that is broken by waves provides a convenient definition of the marginal ice zone for ice/ocean models, 
as the relatively small floes sizes are likely to affect the dynamic and thermodynamic properties of the ice cover \citep{Wiletal13a,Wiletal13b}.

The waves themselves are affected by interactions with the floes.
In particular, wave energy attenuates approximately exponentially with distance into the ice-covered ocean \citep[e.g.][]{Squ&Mor80}.
There is also evidence that the directional wave spectrum becomes isotropic in the ice-covered ocean \citep{wadhams1988attenuation}.
Wave scattering by floes can explain both exponential attenuation and 
spreading of the directional spectrum. 
However, comparisons between numerical scattering models and field data indicate that scattering cannot
account for all of the attenuation experienced by waves \citep{kohout2008elastic,bennetts2010three,Ben&Squ12b}.
Dissipative processes, such as overwash (the wave running over the top of floes), 
floe-floe collisions, and viscoelasticity, must also be considered. 
Wave-induced drift of ice floes may also affect the strength of the attenuation produced.

\cite{Dumetal11} and \cite{Wiletal13a,Wiletal13b} recently proposed a model of wave energy transport in the ice-covered ocean and concomitant wave-induced floe breaking. 
\cite{Mas&LeB89} and \cite{jgrrealism} proposed similar models of wave energy transport in the ice-covered ocean but without floe breaking over 15 years earlier.
The wave energy transport models are based on a modified version of the energy balance equation \citep{komen94},
which includes a source term for wave-ice interactions, $S_{ice}$.
The wave-ice term parametrizes directional scattering and dissipation of energy due to the presence of ice cover.
Source terms that exist in the open ocean, i.e.\ wind input $S_{in}$ and dissipation $S_{ds}$, 
are also modified in the ice-covered ocean, 
e.g.\  by scaling the terms according to the proportion of open ocean present
\citep{Mas&LeB89,Per&Hu96a}.
It is not yet clear how the nonlinear interaction term, $S_{nl}$, should be modified.
The resulting balance equation reads as follows:

\begin{equation}
\frac{\partial E\left(f,\vartheta \right)}{\partial t} + \vec{C}_g\cdot \nabla E\left(f,\vartheta\right) 
=\left(1-f_i\right)\left( S_{in}+S_{ds}\right)+S_{nl}+S_{ice},
\label{balanceeq}
\end{equation}
where $E\left(\tau,\vartheta \right)$ is the wave energy spectrum, as a function of frequency and direction ($f$ and $\vartheta$, respectively); $\vec{C_g}$ is the group velocity; and $f_i$ is the fraction of ice coverage. 

In its most simple form, the wave-ice term is $S_{ice}=-f_{i}\alpha E/d$, 
where $f_{i}\alpha/d$ is the exponential attenuation rate of wave energy per metre due to ice cover.
The quantity $\alpha$ is the attenuation coefficient, and $d$ is the average diameter of the floes.
Using the model of \cite{Ben&Squ12a}, the attenuation rate is related to the proportion of incident wave energy transmitted by a single floe, $Tr$, via
$\alpha=-\log(Tr)$. 
Note that, using this simple form for $S_{ice}$, all reflected wave energy is lost from the system.
Alternatively, conservative, directional scattering can be incorporated, as outlined in \cite{Mas&LeB89}, \cite{jgrrealism} and \citet{ocean_modelling06}.

The attenuation rate is functionally dependent on wave period.
Attenuation rates measured in field experiments
range from approximately $8\times 10^{-4}$\,m$^{-1}$ for 8--9\,s waves, to approximately $2\times 10^{-4}$\,m$^{-1}$ for long waves (periods $>$ 10\,s). 
Although, generally, attenuation rates increase with decreasing wave periods,
a `roll-over' effect, i.e.\ a decrease in attenuation for small wave periods ($<$8\,s) has been noted \citep{wadhams1988attenuation}. 
A summary of recorded attenuation rates is given in \cite{schulz2002spaceborne}. 

Field measurements are normally limited to mild wave conditions. 
Moreover, existing numerical attenuation models are linear.
The functional dependence of the attenuation rate on wave height is, therefore, not properly understood yet.
This includes the onset and subsequent contribution of highly non-linear processes, such as overwash and the drift of the ice floe.

In the present paper, we present an experimental model of wave reflection and transmission, and hence attenuation, by a floe.
The experimental model was implemented in the coastal wave basin at the Coastal Ocean and Sediment Transport (\textsc{coast}) laboratories of Plymouth University, UK. 
The floe is made of a synthetic elastic material, which bends and flexes with the wave motion.
Regular waves of prescribed period and amplitude are generated mechanically.
Wave elevation is measured by arrays of wave gauges before and after the floe.
Despite the idealisations of the real-world phenomenon necessary in a laboratory setting,
experimental facilities allow investigation of transmission in a controlled environment, 
in which the quantities of interest can be measured at high accuracy.

\cite{kohout2007linear} and \cite{montiel2013hydroelastic,montiel2013hydroelasticb} report experimental models of wave-floe interactions,
which are closely related to the experimental model described here. 
However, the previous experiments employed an edge barrier to preclude overwash.
We avoid use of the artificial edge barrier.
Further, we provide evidence that overwash affects reflection and transmission properties of the floe,
using measurements provided by a wave gauge located at the centre of the upper surface of the floe.


\section{Laboratory experiments}
\begin{figure}
  \centering
  \noindent\includegraphics[width=13cm]{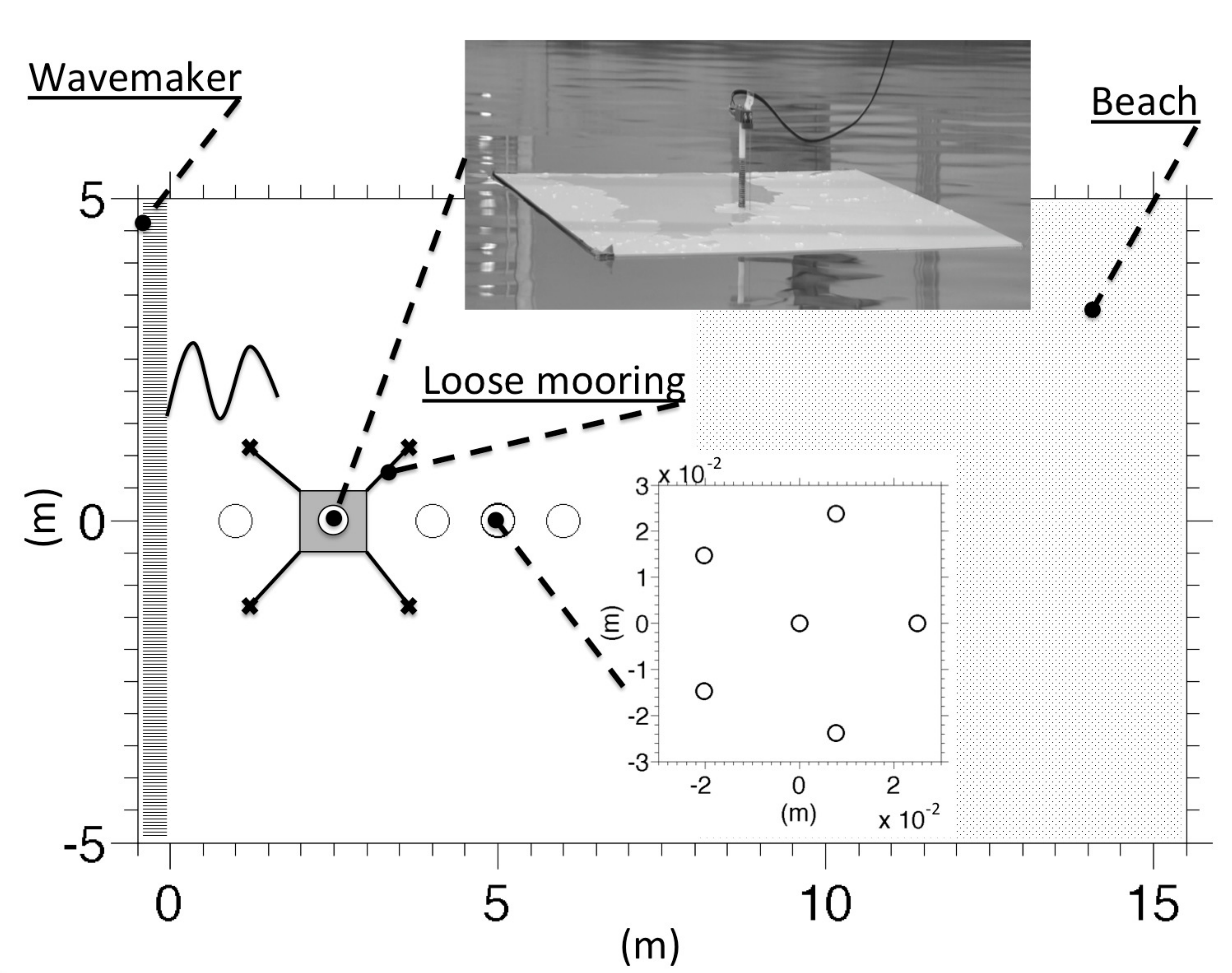}
  \caption{Directional wave basin and experimental set-up.}\label{fig01}
\end{figure}

The laboratory facility consists of a directional wave basin of width of 10\,m, length 15.5\,m and water depth 0.5\,m (see Fig.~\ref{fig01}). 
The tank is equipped with twenty individually controlled active-piston wave makers, which are capable of absorbing incoming waves by measuring the force on the front of the paddle and controlling the velocity \citep{salter1981absorbing}.
At the opposite end, wave energy is dissipated by a beach of slope 1:10. A reflection analysis in the centre of the basin shows that the combined effect of the active pistons and the beach ensures an overall level of reflection lower than 1\% of the incoming energy. Contaminating reflection is filtered from wave records.


At a distance of 2\,m from the wave maker, a plastic sheet was deployed to simulate an ice floe. 
Two different types of plastic were tested: a polypropylene plastic with density of 0.905\,g\,cm$^{-3}$ and Young's modulus 1600\,MPa; and \textsc{pvc} (\textsc{forex}\textregistered) plastic with density 0.500\,g\,cm$^{-3}$ and Young's modulus 500\,MPa. Note that polypropylene has density similar to sea ice but has different rigidity.  
\textsc{Pvc} has a rigidity comparable to sea ice but substantially lower density, which results in a larger freeboard.    
Polypropylene sheets were provided with thicknesses 5\,mm, 10\,mm, 20\,mm and 40\,mm; 
\textsc{pvc} was provided with thicknesses 5\,mm, 10\,mm and 19\,mm. 
The sheets were cut into square floes with side lengths $L_{plate}=1$\,m. 
The experimental set-up was designed to represent full scale wave-ice interactions in the ratio 1:100.

At the wave maker, waves were generated by imposing three different wave periods, namely $T=0.6$\,s, 0.8\,s, and 1\,s, with corresponding wavelengths
$L_{wave}=$0.56\,m, 1\,m and 1.56\,m, respectively. 
The wave fields therefore tested conditions in which the waves were shorter than, equal to and longer than the floe. 
The wave amplitude, $a$, was selected so that the wave steepness $ka$, where $k$ is the wavenumber, 
matched the following values: 0.04, 0.08, 0.1 and 0.15. 
This range includes gently sloping waves ($ka=0.04$ and $0.08$) and storm-like waves ($ka = 0.1$ and 0.15), without reaching the breaking limit \citep[cf.,][]{toffoli10}. 

\begin{figure}
  \centering
  \noindent\includegraphics[width=14cm]{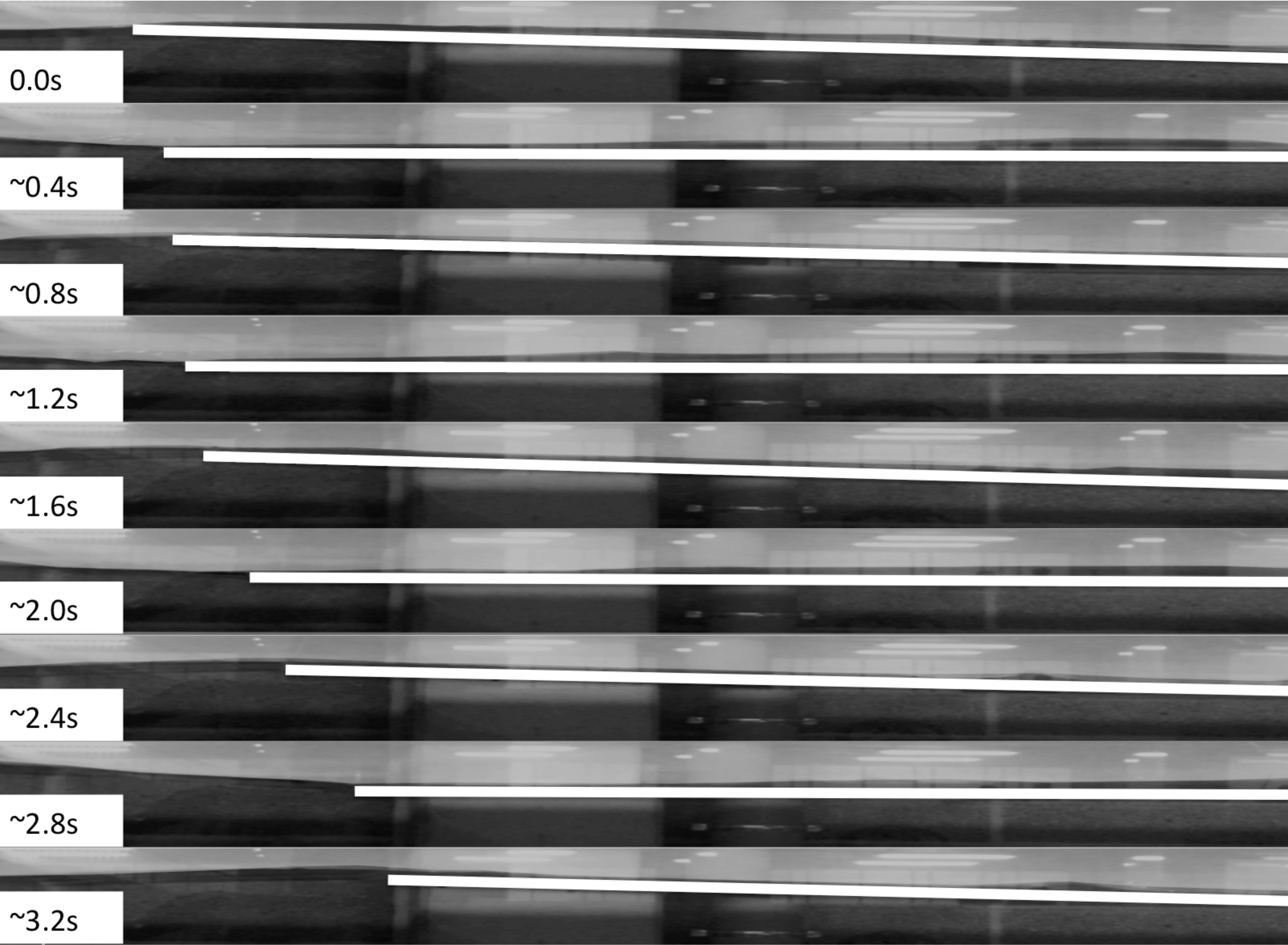}
  \caption{Drift of a floe (polypropylene) under the influence of a monochromatic wave with period $T=0.8s$ and steepness $ka=0.1$.}\label{drift}
\end{figure}

Preliminary tests were conducted to monitor the drift of the floe in a wave flume equipped with glass walls. Only polypropylene plastic was used.
Fig.~\ref{drift} shows the drift in a sequence of images, for an incident wave with period $T=0.8s$ and  steepness $ka=0.1$.
Morrison's equation predicts the floe will obey an almost period motion with a mean forward movement similar to the Stokes drift ($\approx0.001$\,m/s)  \citep{chaplin1984}. 
However, the plastic sheet was only subjected to a forward drift (i.e. no oscillatory components) with speed two orders of magnitude greater ($\approx 0.1$\,m/s) than the Stokes drift. 

In the wave basin experiments, a loose mooring was applied at the four corners of the floe to suppress drift and subsequently avoid collisions with the wave probes. 
The mooring allowed the floe to respond to wave forcing in its six rigid-body degrees of freedom (heave, surge, sway, pitch roll and yaw) and elastic motions. 
Under wave forcing the mooring moved rapidly from being loose to being in tension, due to the drift.
It is worth mentioning that, despite an incipient mooring-free motion, floe kinematics and concurrent wave attenuation might have been partially affected by mooring forces, especially during more severe wave conditions.    

\begin{figure}
  \centering
  \noindent\includegraphics[width=17cm]{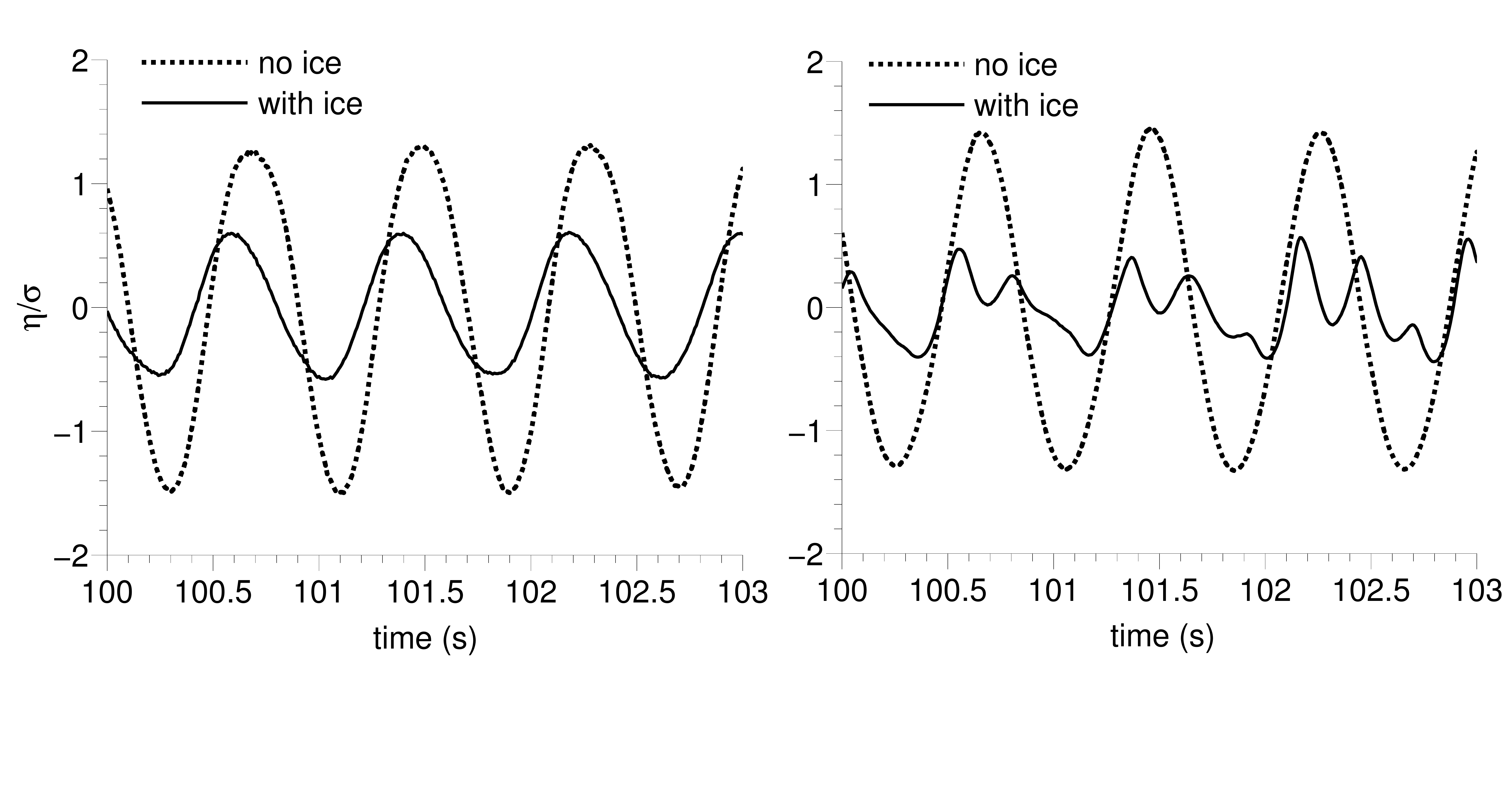}
  \caption{Normalised surface elevation in the lee of a polypropylene plastic sheet of 5\,mm for a wave filed with period $T=0.8$\,s (wavelength 1\,m) and steepness $ka=0.15$.}\label{elevation}
\end{figure} 

The water surface elevation, $\eta$, was monitored with capacitance gauges at a sampling frequency of 128\,Hz. 
One gauge was deployed approximately 1\,m in front of the floe to capture the incident and reflected waves. 
In the lee of the floe, three probes were deployed every metre to track the evolution of the transmitted wave field. 
At 2\,m from the rear edge of the floe a six-probe array, arranged as a pentagon of radius of 0.25\,m and a middle probe, 
was deployed to monitor the directional properties of the wave field. 
In order to quantify the depth of overwash (and concurrent wave motion), 
a mini-gauge was deployed in the middle of the upper surface of the floe (see photo in Fig.~\ref{fig01}). Its weight was a negligible small fraction of the floe's weight. 
For each configuration, five-minute time series were recorded. 

\section{Results}

\subsection{Water surface elevation} \label{surf}

Fig.~\ref{elevation} shows an example of time series for the surface elevation provided by a wave gauge in the lee of the floe for gently sloping waves ($ka=0.04$, left panel) and storm-like waves ($ka=0.15$, right panel).
The corresponding time series from a control test, conducted without the floe, is also shown. 
For the gently sloping incident waves a notable attenuation in wave energy is detected. The transmitted waves, however, retain a regular profile. 
An even more substantial attenuation of the transmitted waves is recorded for the storm-like incident waves. The wave profile assumes a more irregular shape as a result of a far more complicated propagation of energetic waves. In particular, waves overwash the floe. 

\begin{figure}
 \centering
  \noindent\includegraphics[width=15cm]{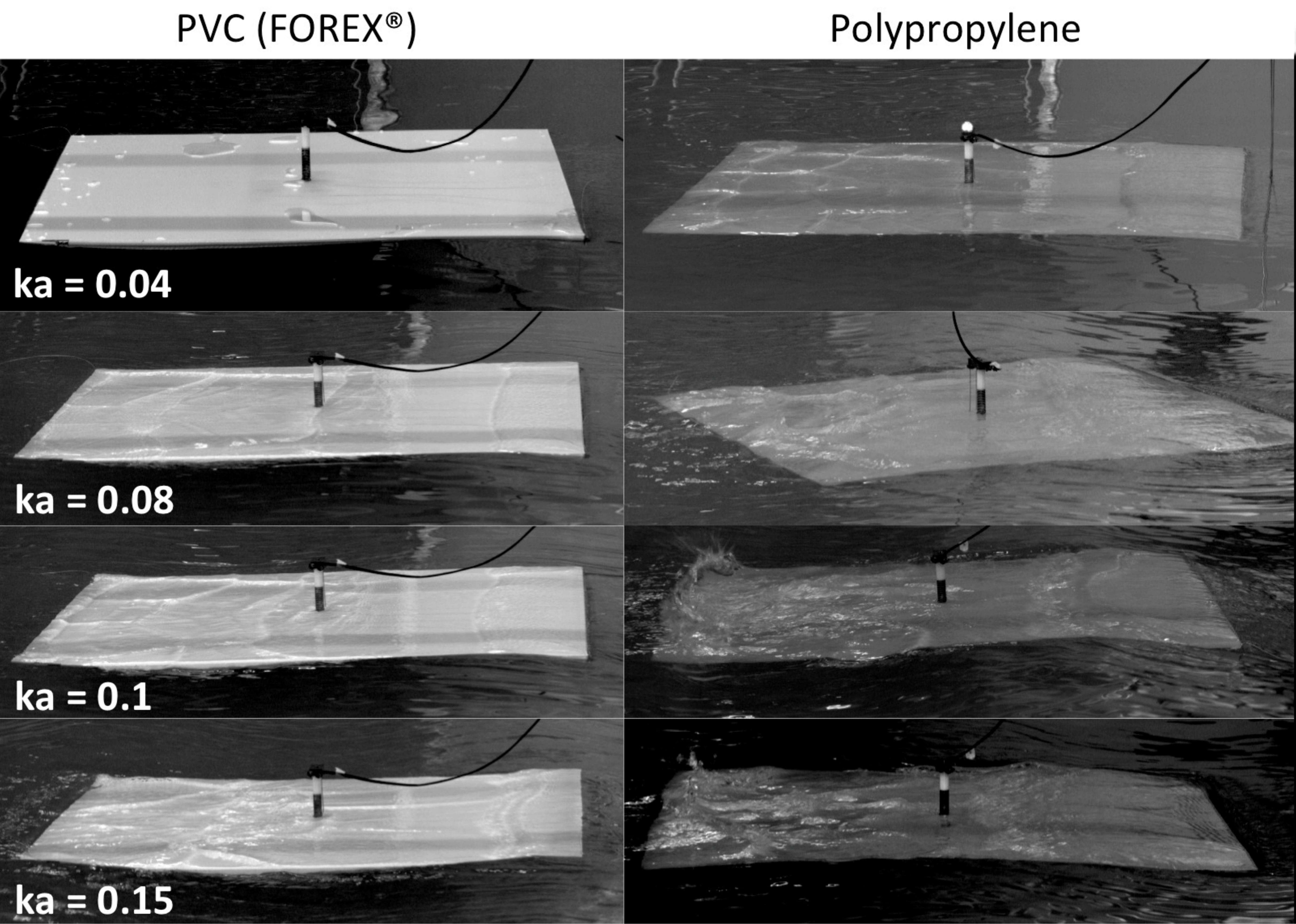}  
  \caption{Overwash of a polypropylene floe (right panels) and \textsc{pvc} floe (left panels) for a wave field with wavelength equal to the floe length and thickness of 10\,mm.}\label{photos}
\end{figure}

Fig.~\ref{photos} shows examples of overwash behaviours for different wave steepnesses
and for the two types of plastic used. 
The floe is submerged by the wave, either partially (for \textsc{pvc}, which has a larger freeboard and/or gentle waves) 
or fully (for polypropylene and energetic waves). 

Shallow-water waves propagate in the overwashed fluid. 
These waves are generated at both ends of the floe and interact with each other. Therefore, wave fields may substantially steepen in the overwashed region. 
This often results in wave breaking and hence energy dissipation, thus contributing to attenuation and generation of spurious high frequency free wave components. 
Moreover, steep waves intensify the impact of the floe onto the water surface (slamming). 
This induces stronger loads on the structure, and contributes to generating spurious high frequency (free) wave components. 
We attribute
the irregularity observed in the lee of the floe for steep waves
to a combination of slamming and overwash leaving the floe's edge.

\begin{figure}
  \centering
  \noindent\includegraphics[width=17cm]{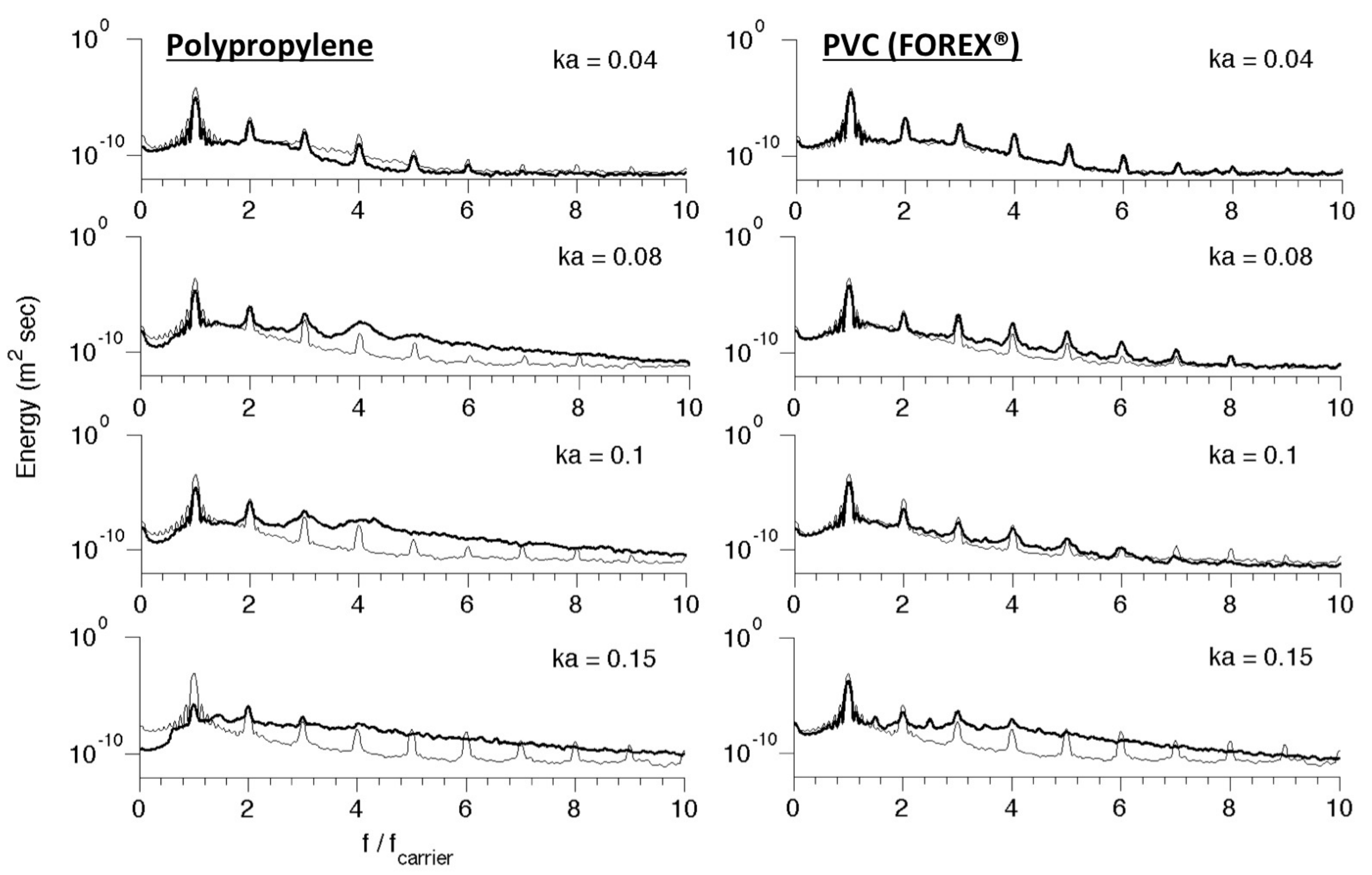}
  \caption{Variation of the incident wave spectrum due to wave-ice interaction: incident wave field (thin solid line); transmitted wave (thick solid line). Wave field characterised by a period of 0.8$s$ (i.e. wavelength of 1\,m).}\label{spec}
\end{figure}

Fig.~\ref{spec} shows incident (from control tests) and transmitted (in the lee of the floe) wave energy spectra.
Spectra are computed from segments of 4096 consecutive records and averaged over the entire time series. Note that the incident wave is characterised by the dominant component (monochromatic wave generated at the wave maker) and  higher harmonics, i.e. phase-locked bound waves. 

For the smallest wave steepness tested ($ka=0.04$), the transmitted wave energy spectrum of the  \textsc{pvc} floe is very close to that of the incident wave.
Slight attenuation of wave energy is detected for polypropylene floe. 
Overwash and slamming do not occur for the mild conditions. 

Attenuation of the dominant spectral component increases, as incident steepness increases.
This is particularly evident for the polypropylene floe, which is subjected to intense overwash and slamming.
The high frequency tails of the spectra for the Polyproplene floe, contain greater energy than the incident spectra for the three largest steepnesses.
Further, the harmonics are smeared out. This
substantiates a relationship with the intensification of overwash and slamming. 
Consistent with this hypothesis, growth of the upper tail is far weaker for \textsc{pvc} floes.

\begin{figure}
  \centering
  \noindent\includegraphics[width=17cm]{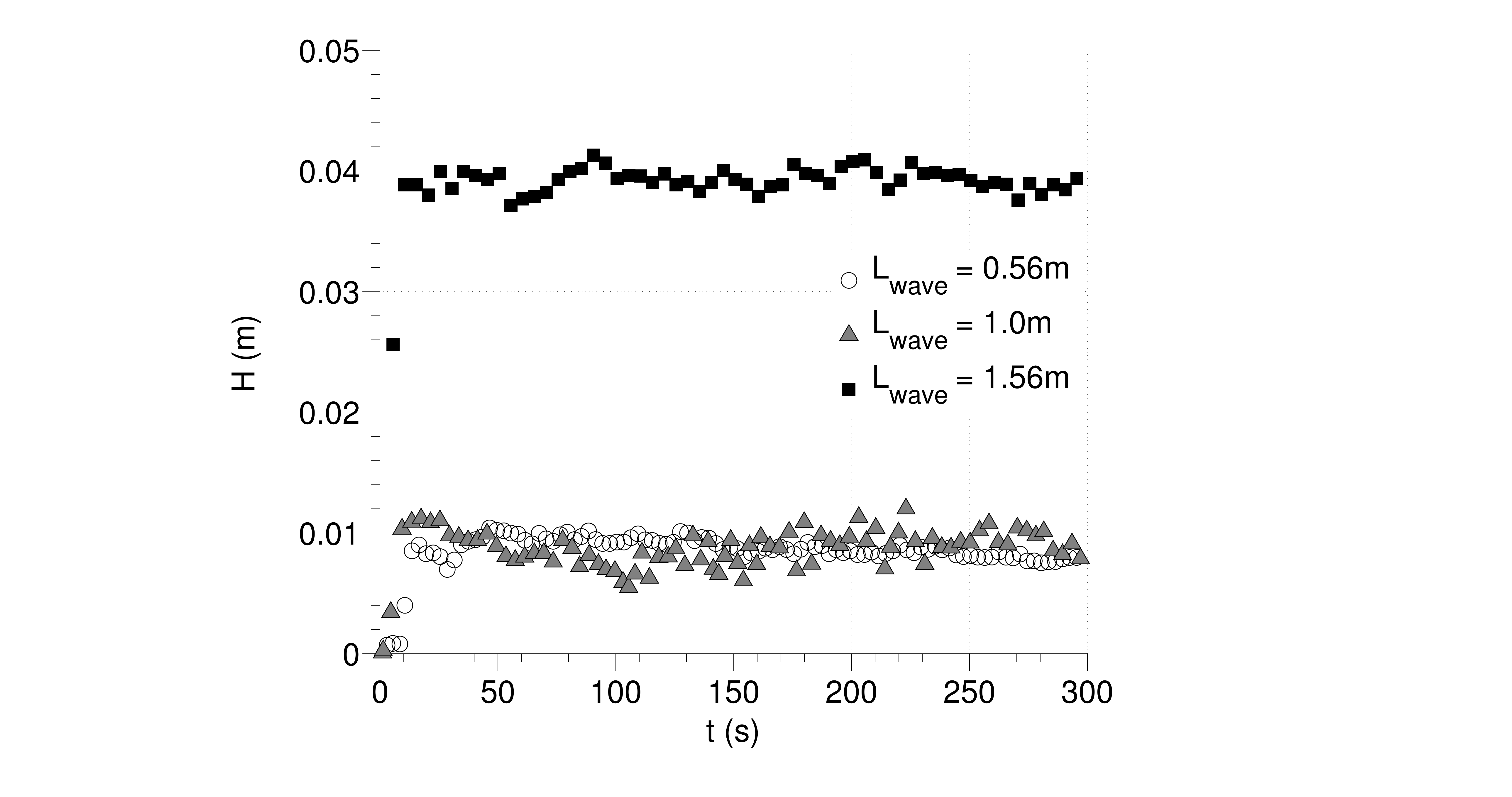}
  \caption{Temporal variation of zero down-crossing wave height: Polypropylene floe.}\label{heightP}
\end{figure}

\begin{figure}
  \centering
  \noindent\includegraphics[width=17cm]{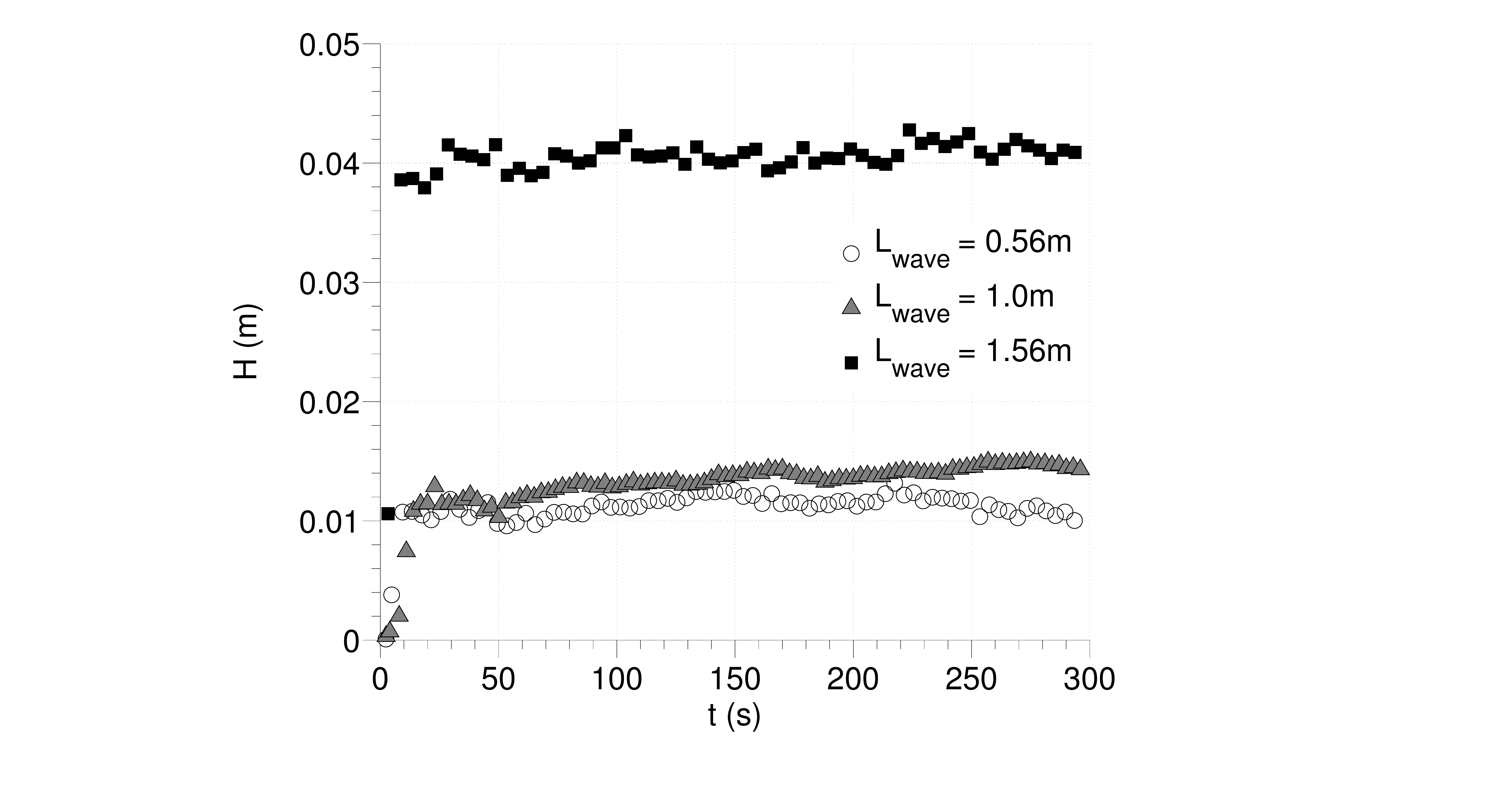}
  \caption{Temporal variation of zero down-crossing wave height: \textsc{Pvc} floe.}\label{heightF}
\end{figure}

Figs.~\ref{heightP} and \ref{heightF} show example time series of wave heights, for the case $ka=0.1$.
Here, wave heights are extracted from records using a standard zero down-crossing method.
No pattern is visible, as the mooring transitions from being loose to being engaged, besides the initial run-up as waves are generated.
Therefore, although we note that
mooring forces may contribute to the generation of high frequency components too (by affecting the way the floe slams onto the water surface), there is no evidence of this in our measurements.
Similar behaviour is observed for lower and higher steepness conditions (not shown).

\subsection{Reflection and transmission} 

\begin{figure}
  \centering
  \noindent\includegraphics[width=12cm]{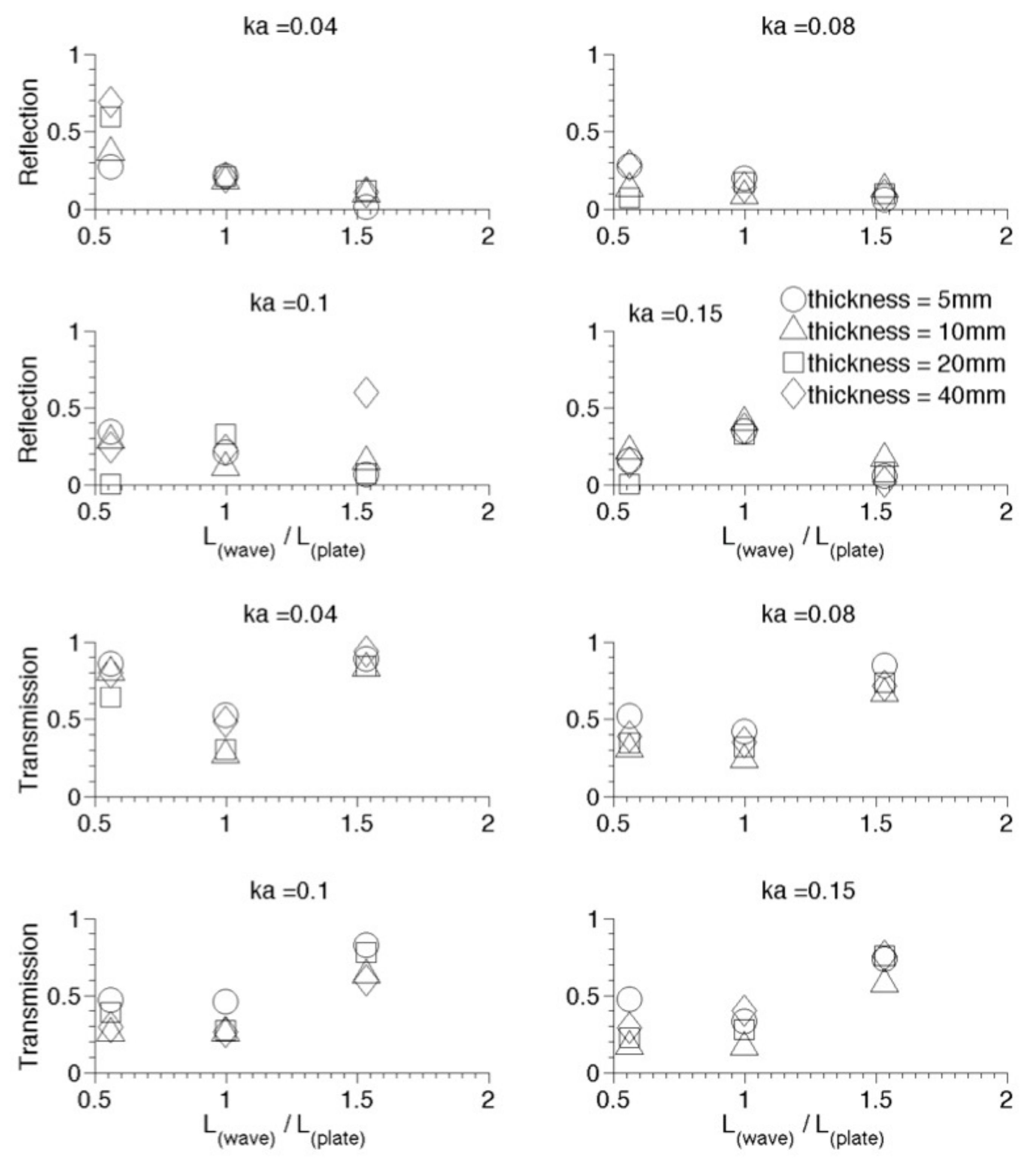}
  \caption{Average transmission coefficient ($T$) against the normalised wavelength as a result of the interaction with a polypropylene floe: thickness = 5\,mm (o); thickness = 10\,mm ($\triangle$); thickness = 20\,mm ($\Box$); and thickness = 40\,mm ($\Diamond$).}\label{TrvsL_ppl}
\end{figure}

\begin{figure}
  \centering
  \noindent\includegraphics[width=12cm]{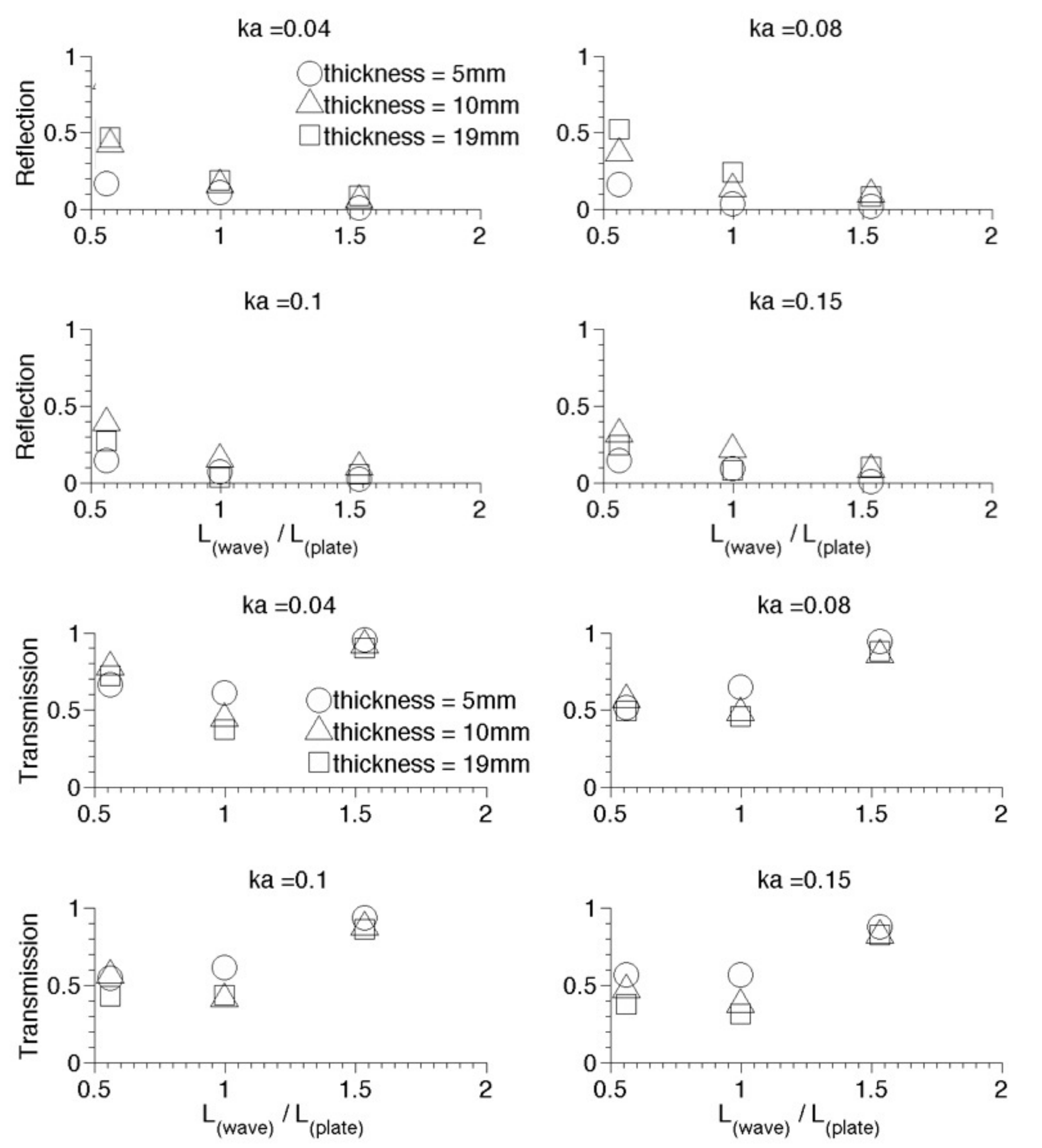}
  \caption{Average transmission coefficient ($T$) against the normalised wavelength as a result of the interaction with a \textsc{pvc} floe: thickness = 5\,mm (o); thickness = 10\,mm ($\triangle$); thickness = 19\,mm ($\Box$).}\label{TrvsL_forex}
\end{figure}

Due to the irregular behaviour of the transmitted waves for storm-like incident waves, we quantify
reflection and transmission in terms of average wave height. The latter is extracted from the records with a standard zero down-crossing analysis and further averaged over the entire time series.
Let the average wave height of the incident wave, from the control tests, be denoted $H_{i}$;
the average wave height in front of the floe is denoted $H_{front}$; and in the lee of the floe is denoted $H_{rear}$. 
Reflection and transmission coefficients are computed as $R = (H_{front} / H_{i}) - 1$ and $T = H_{rear} /  H_i$, respectively. 
Note that this approach contains both linear and nonlinear contributions to the wave field. We also remark that reflection from the beach and the piston wave maker are removed. Therefore, $R$ only refers to the floe-induced reflection.

Average reflection and transmission coefficients, as functions of a normalised wavelength (i.e.\ a ratio of wavelength to floe length), 
for the different wave steepness and floe thickness are presented in Figs.~\ref{TrvsL_ppl} and \ref{TrvsL_forex} (polyporpylene and \textsc{pvc}, respectively). 
Significant variation of reflection and transmission with wavelength, thickness and steepness are evident.

At the front edge of the floe, a portion of the incident wave is back-scattered. 
For low wave steepnesses, reflection increases with increasing floe thickness,
in agreement with predictions of linear models \citep[e.g.][]{meylan_squire94}.
However, the relationship between reflection and thickness becomes more complicated for larger steepnesses.

Decreasing reflection as wavelength increases is also apparent for low steepness waves.
This is also consistent with numerical models \citep[e.g.][]{lavrenov2000chapter}. 
The relationship becomes more complicated for larger steepnesses although it is approximately maintained for all steepnesses by the more compliant \textsc{pvc} floes.

It would be reasonable to expected that, for low wave steepnesses, the transmission coefficient would follow a complementary monotonic increasing trend with increasing wavelength. 
However, the experimental data do not show this expected behaviour.
Instead, transmission is minimum for a wavelength equal to the floe length.
This resembles the roll-over in attenuation found in field data, although we do not claim that our findings are responsible for the roll-over phenomenon.

As waves become more energetic (that is to say, wave steepness increases), wave reflection generally reduces. 
The difference is most pronounced for shorter waves. 
Strikingly, reductions in reflection for larger steepnesses are matched by reductions in transmission. 
The decrease in transmission is also most significant for shorter waves. 
This results in a qualitative change in the dependence of transmission on wavelength, so that the `roll-over' effect is diminished. 
For $ka \geq 0.08$, transmission is approximately constant for $L_{wave} / L_{plate} \leq 1$, 
and then increases for the longest wave 
(see bottom panels in Figs.~\ref{TrvsL_ppl} and \ref{TrvsL_forex}). 
Despite some scatter, transmission trends are the same for the polypropylene and \textsc{pvc} floes. 
The more compliant \textsc{pvc} floes do, however, transmit a larger proportion of the waves. 


\subsection{Effect of overwash}

\begin{figure}
  \centering
  \noindent\includegraphics[width=14cm]{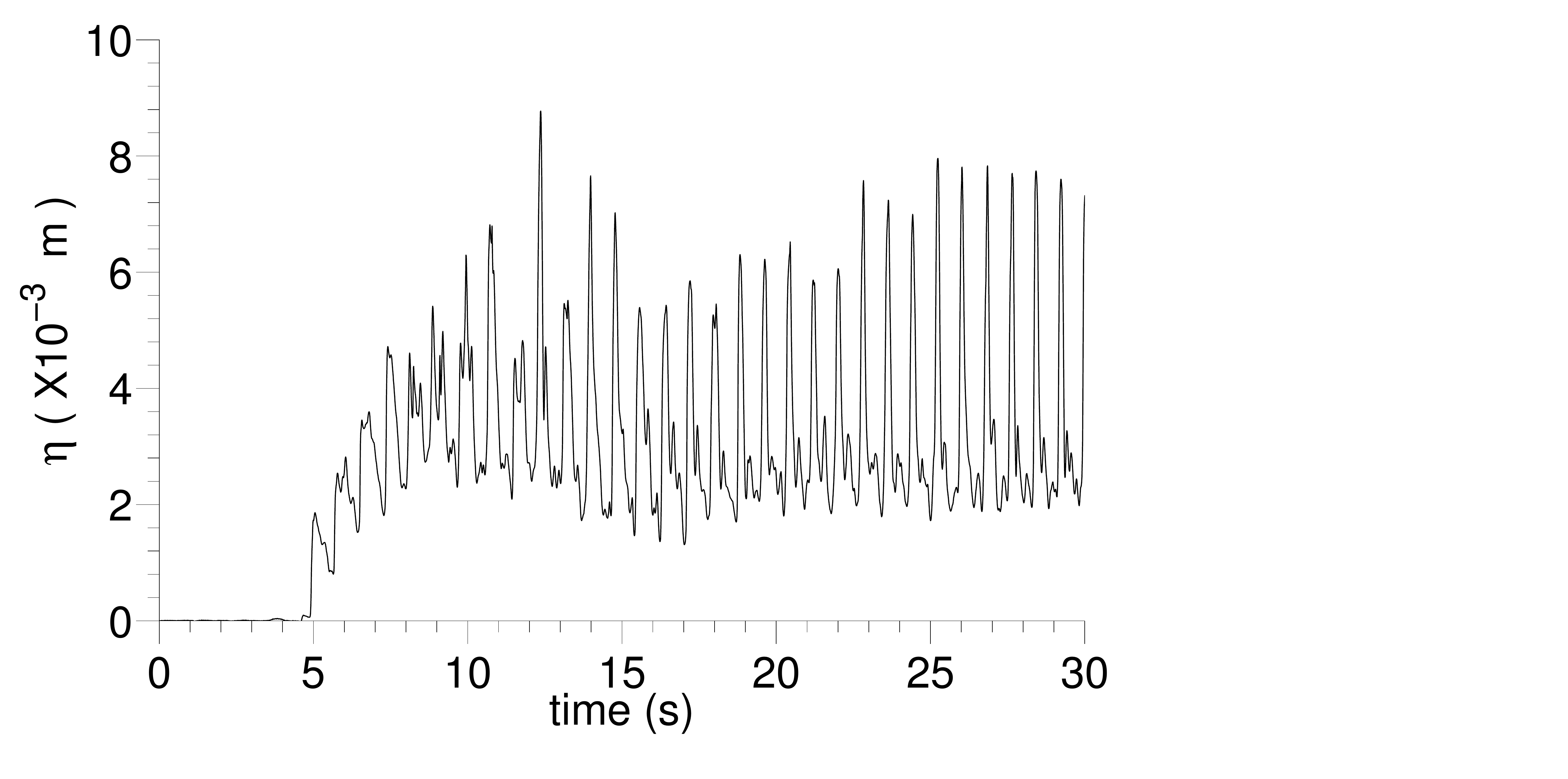} 
  \caption{Example of overwash water surface elevation on a polypropylene plastic sheet of thickness 10\,mm in a wave field of wavelength 1\,m and steepness $ka=0.15$.}\label{overwash}
\end{figure}

\begin{figure}
  \centering
  \noindent\includegraphics[width=14cm]{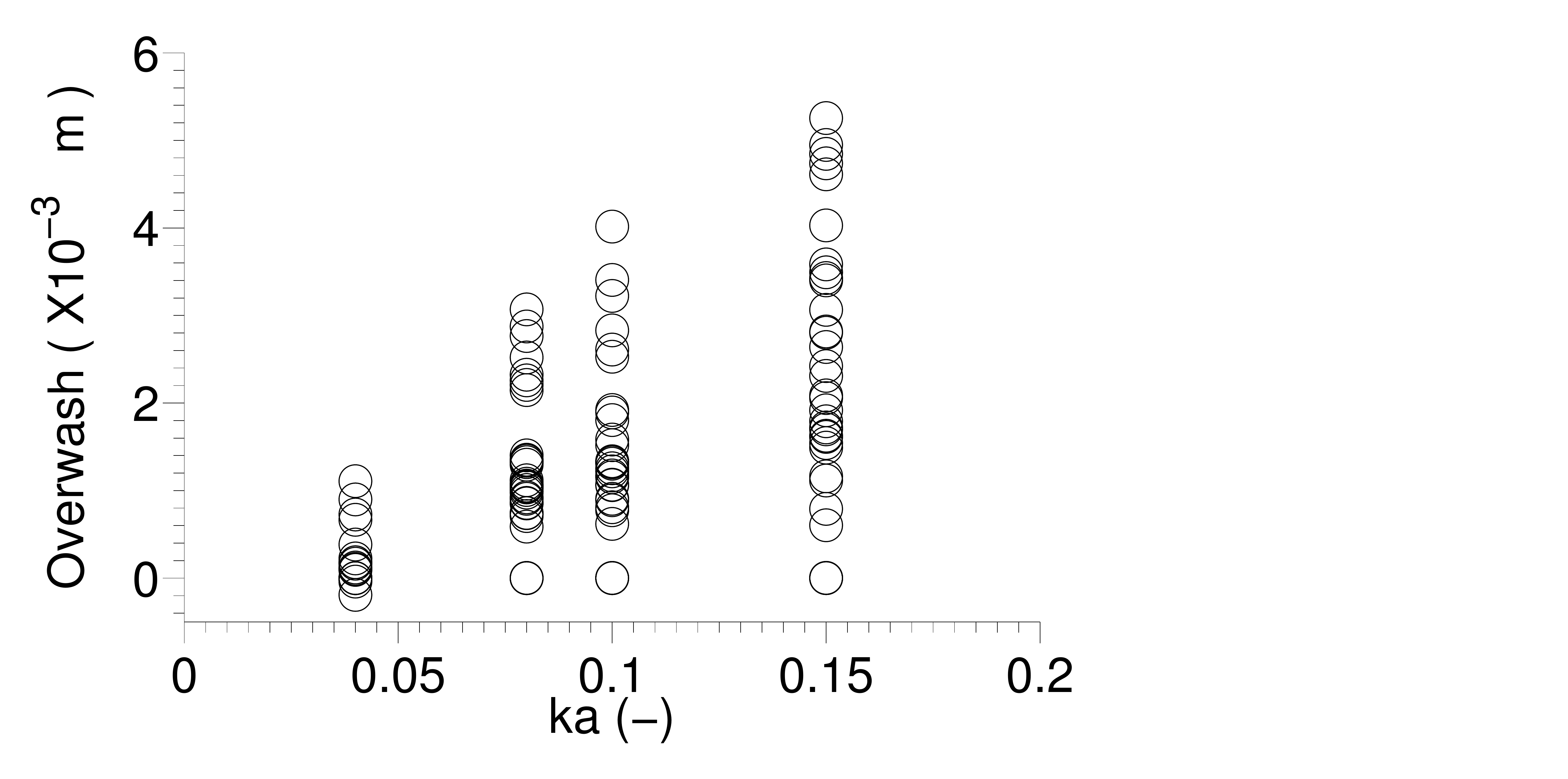}   
  \caption{Overwash depth as a function of wave steepness.}\label{overwash_and_steepness}
\end{figure}

\begin{figure}
  \centering
  \noindent\includegraphics[width=13cm]{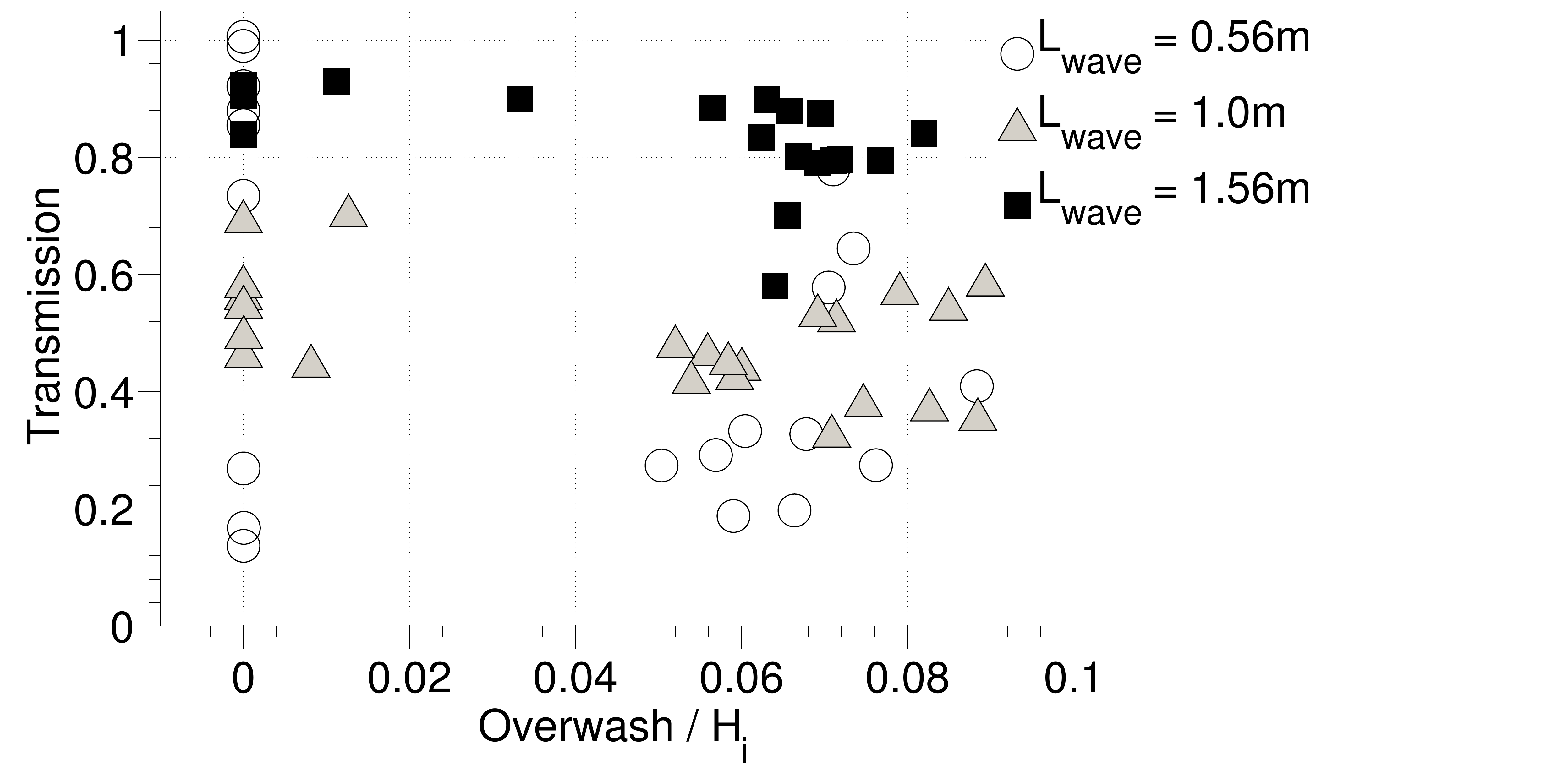}
  \caption{Transmission as a function of normalised overwash depth for polypropylene floes.}\label{TvsOvpoly}
\end{figure}

\begin{figure}
  \centering
  \noindent\includegraphics[width=13cm]{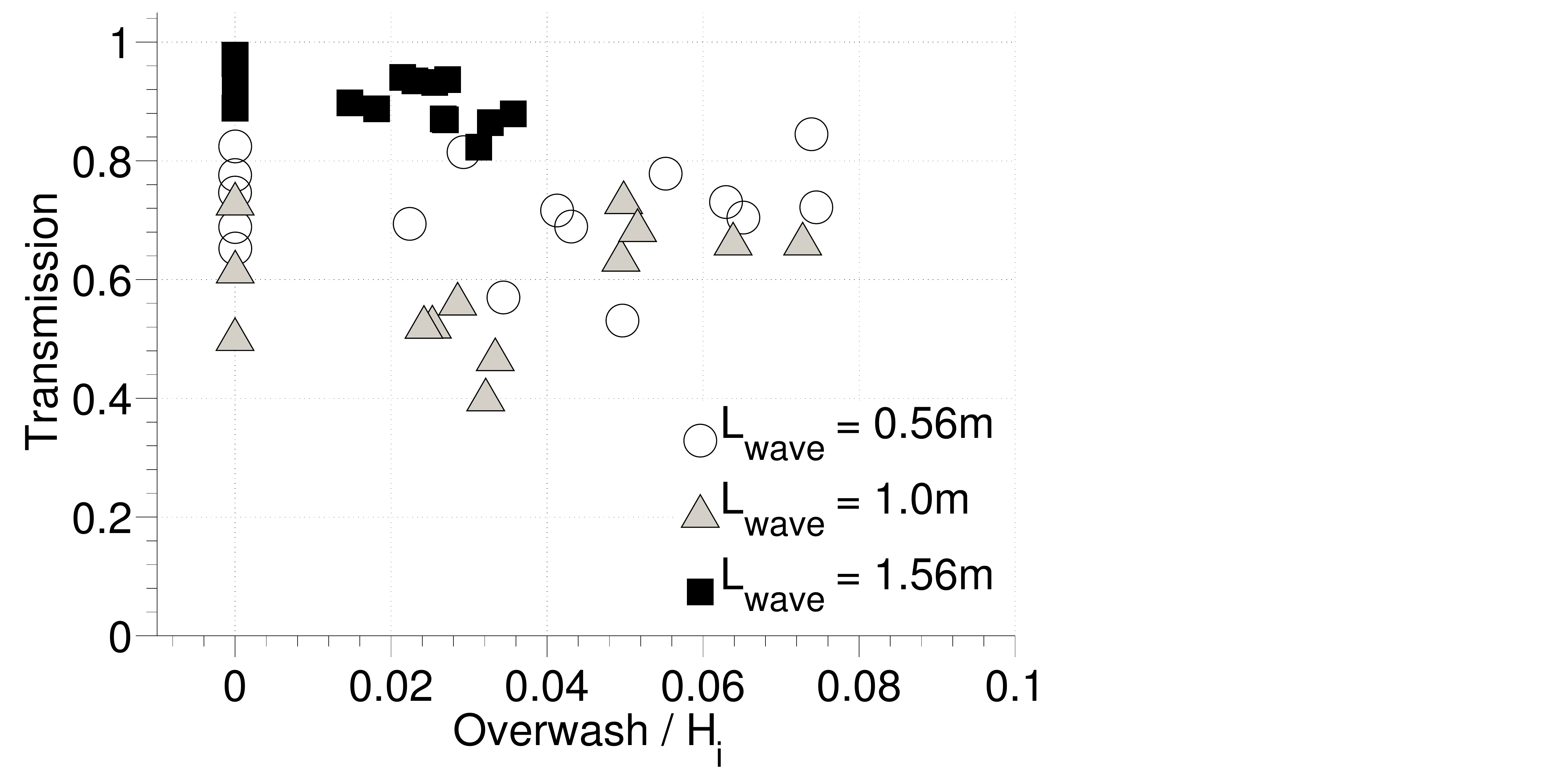}
  \caption{As in Fig.~\ref{TvsOvpoly} but for \textsc{pvc} floes.}\label{TvsOvpvc}
\end{figure}

\begin{figure}
   \centering
  \noindent\includegraphics[width=14cm]{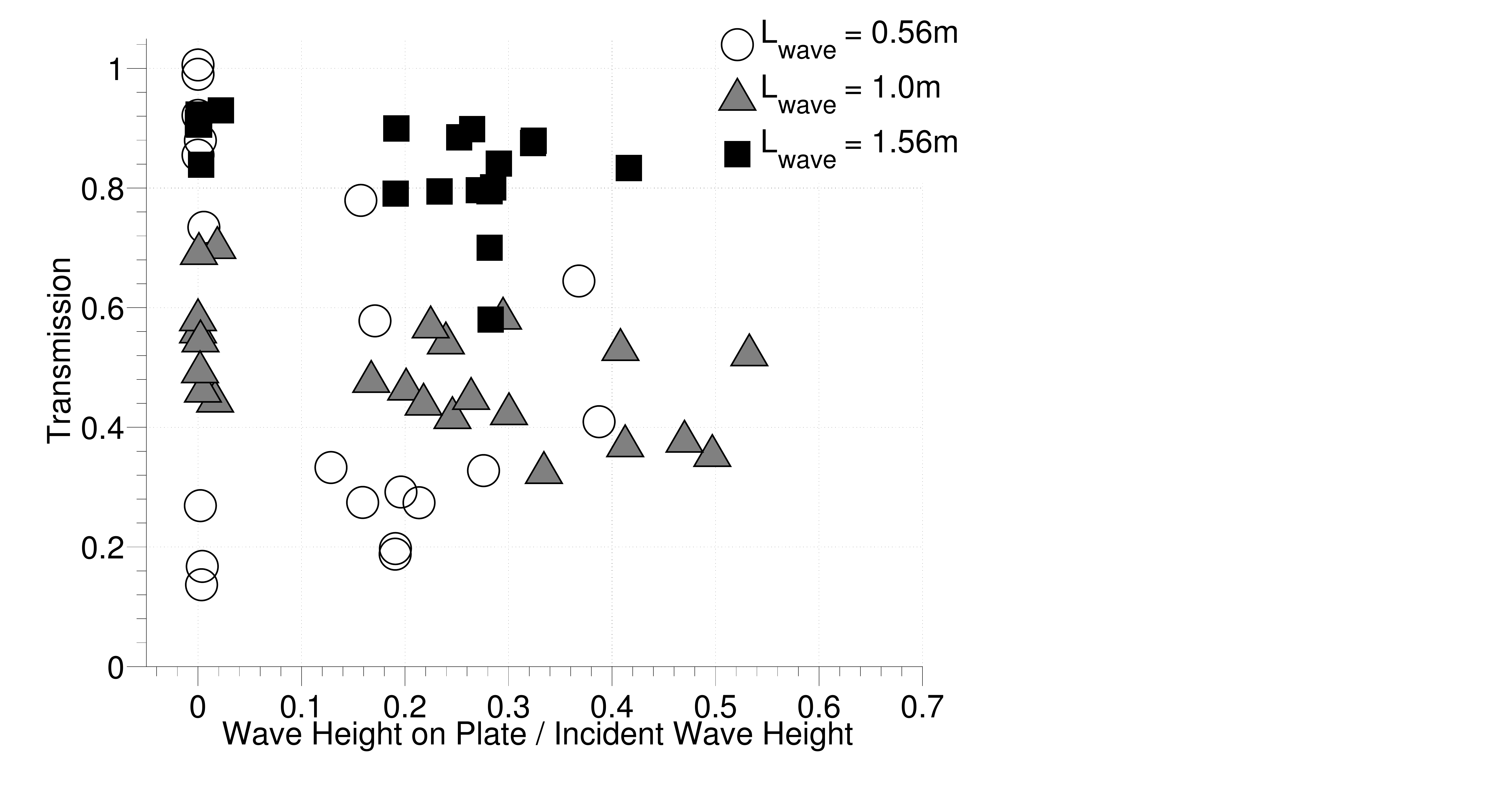}
  \caption{Transmission as a function of normalised overwash significant wave height for polypropylene floes.}\label{TvsWpoly}
\end{figure}

\begin{figure}
   \centering
  \noindent\includegraphics[width=14cm]{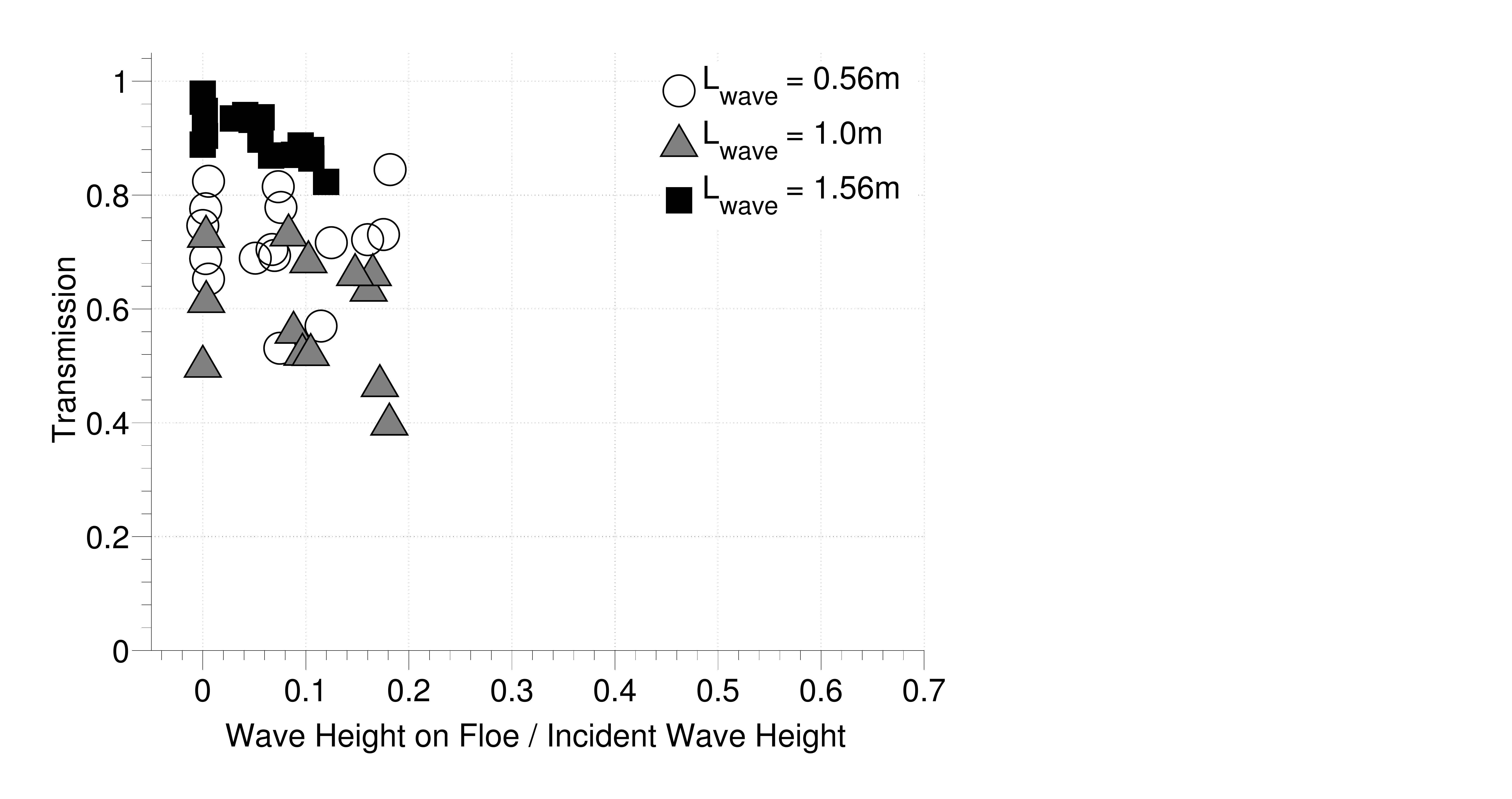}
  \caption{As in Fig.~\ref{TvsWpoly} but for \textsc{pvc} floes.}\label{TvsWpvc}
\end{figure}

Low steepness waves gently propagate through the floe and hence attenuation is attributed to wave reflection and any energy lost during the flexure of the floe. With the occurrence of overwash, however, part of the wave energy propagates directly over the floe, and, thus, is no longer reflected by the front edge of the sheet. 
This helps to explain the slight reduction in the reflection coefficient observed for more energetic wave fields. 
Dissipation of wave energy in the overwashed fluid helps explain 
the reduction of transmission observed for more energetic waves.


The depth of overwashed fluid and, hence, wave motion over the floe was measured by a wave gauge. 
Fig.~\ref{overwash} shows an example time series provided by the wave gauge. 
From here on, overwash depth refers to the time series mean. 
Dependence of overwash depth on steepness is confirmed in Fig.~\ref{overwash_and_steepness}. 

The relationship between transmission and overwash depth is shown in Figs.~\ref{TvsOvpoly} and \ref{TvsOvpvc},
for the polypropylene and \textsc{pvc} floes, respectively.
Overwash depth is normalised by the incident wave height. 
It is evident that the mechanical properties of the floe affect overwash depth. 
In particular, overwash depth is larger for the polypropylene floes, which have smaller freeboards than the \textsc{pvc} floes.

An overall trend of reducing transmission with increasing overwash depth is evident.  
The relationship is mostly clear for the polypropylene floes.
Monotonic reduction of transmission only occurs for the longest waves ($L_{wave} = 1.56\,m$). 
For shorter waves, there is normally a sudden drop of the transmission when normalised overwash depth exceeds approximately 0.02. 
Transmission then further decreases with increasing overwash depth, albeit at a very small rate for polypropylene floes. 
In comparison, for the \textsc{pvc} floes there is some evidence of overturning, leading to increasing transmission, for deeper overwash. 
Note that, eventually, transmission seems to level off for the largest overwash depths.   

The correlation between transmission and normalised overwash depth is weak.
Overwash depth only estimates a uniform layer of water over the floe and it does not consider the intensity of wave activity. 
We therefore consider transmission as a function of the significant wave height of the overwashed fluid (normalised with respect to the incident wave height) in Figs.~\ref{TvsWpoly} and \ref{TvsWpvc}. 
The relationship between transmission and overwash significant wave height  follows a more consistent trend. 
In particular, there is a monotonic decrease of transmission with increasing significant wave height. 
The rate of decrease is, largely, insensitive to wavelength, although data are more scattered for shorter waves.

\section{Conclusions}

An experimental model of reflection and transmission of ocean waves by a solitary ice floe has been reported.
The experimental tests were conducted in the coastal directional wave basin of Plymouth University,
using regular incident waves with different wave periods, amplitudes and steepnesses. 
Wave fields were selected to range from mild to storm-like conditions.
The floe was modelled by a square plastic sheet. 
Two different plastics were used, and different thicknesses were tested. 
The wave elevation in front of the floe and in its lee was measured by an array of wave gauges. Only the overall amount of wave energy was analysed to estimate reflection and transmission coefficients. 
Wave overwash of the floe was measured by a mini wave gauge deployed in the centre of the upper surface of the floe. 

The following key conclusions were drawn from the data:
\begin{enumerate}
  \item Transmitted waves retain regularity for small-steepness incident waves. However, transmitted waves are highly irregular for large-deepness incident waves. This was attributed to wave overwash of the floe and slamming of the floe by energetic waves.
  
  \item Wave reflection in the experiments is qualitatively consistent with linear numerical model for low wave steepnesses, i.e.\ reflection increases for thicker floes, and decreases for longer waves. However, these relationships are not maintained for large steepnesses.
  
  \item Transmission cannot be inferred from reflection. 
  Even for the most gentle wave steepness, the transmission is not monotonic with wavelength, instead taking a minimum when the wavelength is equal to the floe length.
  
  \item In general, overwash increases as wave steepness increases.
Transmission decreases slightly with increased depth of the overwashed fluid (normalised with respect to the incident wave height), 
and more significantly with the significant wave height of the waves in the overwashed fluid. 


\end{enumerate}



\section{Acknowledgements}
Experiments were supported by the Small Research Grant Scheme of the School of Marine Science and Engineering of Plymouth University and performed when AT and AA were appointed at Plymouth University.
LB acknowledges funding support from the Australian Research Council (DE130101571) and the Australian Antarctic Science Grant Program (Project 4123).


\end{document}